\documentclass[conference]{IEEEtran}
\IEEEoverridecommandlockouts

\usepackage{cite}
\ifCLASSINFOpdf
  \usepackage[pdftex]{graphicx}
\else
  \usepackage[dvips]{graphicx}
\fi
\usepackage{amsmath}
\usepackage{algorithmic}
\usepackage{array}
\usepackage{tabularx}
\usepackage{url}
\usepackage{xcolor}
\usepackage{booktabs,amsfonts}
\usepackage[nolist,nohyperlinks]{acronym}
\usepackage{subfigure}
\usepackage{microtype}

\usepackage{amssymb}
\usepackage{amsfonts}
\usepackage{mathrsfs}
\usepackage{xspace}
\usepackage{bm}
\usepackage{upgreek}

\newcommand{\safemath}[2]{\newcommand{#1}{\ensuremath{#2}\xspace}}

\safemath{\bma}{\mathbf{a}}
\safemath{\bmb}{\mathbf{b}}
\safemath{\bmc}{\mathbf{c}}
\safemath{\bmd}{\mathbf{d}}
\safemath{\bme}{\mathbf{e}}
\safemath{\bmf}{\mathbf{f}}
\safemath{\bmg}{\mathbf{g}}
\safemath{\bmh}{\mathbf{h}}
\safemath{\bmi}{\mathbf{i}}
\safemath{\bmj}{\mathbf{j}}
\safemath{\bmk}{\mathbf{k}}
\safemath{\bml}{\mathbf{l}}
\safemath{\bmm}{\mathbf{m}}
\safemath{\bmn}{\mathbf{n}}
\safemath{\bmo}{\mathbf{o}}
\safemath{\bmp}{\mathbf{p}}
\safemath{\bmq}{\mathbf{q}}
\safemath{\bmr}{\mathbf{r}}
\safemath{\bms}{\mathbf{s}}
\safemath{\bmt}{\mathbf{t}}
\safemath{\bmu}{\mathbf{u}}
\safemath{\bmv}{\mathbf{v}}
\safemath{\bmw}{\mathbf{w}}
\safemath{\bmx}{\mathbf{x}}
\safemath{\bmy}{\mathbf{y}}
\safemath{\bmz}{\mathbf{z}}
\safemath{\bmzero}{\mathbf{0}}
\safemath{\bmone}{\mathbf{1}}

\bmdefine{\biad}{a}
\bmdefine{\bibd}{b}
\bmdefine{\bicd}{c}
\bmdefine{\bidd}{d}
\bmdefine{\bied}{e}
\bmdefine{\bifd}{f}
\bmdefine{\bigd}{g}
\bmdefine{\bihd}{h}
\bmdefine{\biid}{i}
\bmdefine{\bijd}{j}
\bmdefine{\bikd}{k}
\bmdefine{\bild}{l}
\bmdefine{\bimd}{m}
\bmdefine{\bind}{n}
\bmdefine{\biod}{o}
\bmdefine{\bipd}{p}
\bmdefine{\biqd}{q}
\bmdefine{\bird}{r}
\bmdefine{\bisd}{s}
\bmdefine{\bitd}{t}
\bmdefine{\biud}{u}
\bmdefine{\bivd}{v}
\bmdefine{\biwd}{w}
\bmdefine{\bixd}{x}
\bmdefine{\biyd}{y}
\bmdefine{\bizd}{z}

\bmdefine{\bixid}{\xi}
\bmdefine{\bilambdad}{\lambda}
\bmdefine{\bimud}{\mu}
\bmdefine{\bithetad}{\theta}
\bmdefine{\biphid}{\phi}
\bmdefine{\bideltad}{\delta}

\safemath{\bmia}{\biad}
\safemath{\bmib}{\bibd}
\safemath{\bmic}{\bicd}
\safemath{\bmid}{\bidd}
\safemath{\bmie}{\bied}
\safemath{\bmif}{\bifd}
\safemath{\bmig}{\bigd}
\safemath{\bmih}{\bihd}
\safemath{\bmii}{\biid}
\safemath{\bmij}{\bijd}
\safemath{\bmik}{\bikd}
\safemath{\bmil}{\bild}
\safemath{\bmim}{\bimd}
\safemath{\bmin}{\bind}
\safemath{\bmio}{\biod}
\safemath{\bmip}{\bipd}
\safemath{\bmiq}{\biqd}
\safemath{\bmir}{\bird}
\safemath{\bmis}{\bisd}
\safemath{\bmit}{\bitd}
\safemath{\bmiu}{\biud}
\safemath{\bmiv}{\bivd}
\safemath{\bmiw}{\biwd}
\safemath{\bmix}{\bixd}
\safemath{\bmiy}{\biyd}
\safemath{\bmiz}{\bizd}

\safemath{\bmxi}{\bixid}
\safemath{\bmlambda}{\bilambdad}
\safemath{\bmmu}{\bimud}
\safemath{\bmtheta}{\bithetad}
\safemath{\bmphi}{\biphid}
\safemath{\bmdelta}{\bideltad}

\safemath{\bA}{\mathbf{A}}
\safemath{\bB}{\mathbf{B}}
\safemath{\bC}{\mathbf{C}}
\safemath{\bD}{\mathbf{D}}
\safemath{\bE}{\mathbf{E}}
\safemath{\bF}{\mathbf{F}}
\safemath{\bG}{\mathbf{G}}
\safemath{\bH}{\mathbf{H}}
\safemath{\bI}{\mathbf{I}}
\safemath{\bJ}{\mathbf{J}}
\safemath{\bK}{\mathbf{K}}
\safemath{\bL}{\mathbf{L}}
\safemath{\bM}{\mathbf{M}}
\safemath{\bN}{\mathbf{N}}
\safemath{\bO}{\mathbf{O}}
\safemath{\bP}{\mathbf{P}}
\safemath{\bQ}{\mathbf{Q}}
\safemath{\bR}{\mathbf{R}}
\safemath{\bS}{\mathbf{S}}
\safemath{\bT}{\mathbf{T}}
\safemath{\bU}{\mathbf{U}}
\safemath{\bV}{\mathbf{V}}
\safemath{\bW}{\mathbf{W}}
\safemath{\bX}{\mathbf{X}}
\safemath{\bY}{\mathbf{Y}}
\safemath{\bZ}{\mathbf{Z}}

\safemath{\bZero}{\mathbf{0}}
\safemath{\bOne}{\mathbf{1}}
\safemath{\bDelta}{\mathbf{\Delta}}
\safemath{\bLambda}{\mathbf{\UpLambda}}
\safemath{\bPhi}{\mathbf{\Upphi}}
\safemath{\bSigma}{\mathbf{\Upsigma}}
\safemath{\bOmega}{\mathbf{\Upomega}}
\safemath{\bTheta}{\mathbf{\Uptheta}}

\bmdefine{\biAd}{A}
\bmdefine{\biBd}{B}
\bmdefine{\biCd}{C}
\bmdefine{\biDd}{D}
\bmdefine{\biEd}{E}
\bmdefine{\biFd}{F}
\bmdefine{\biGd}{G}
\bmdefine{\biHd}{H}
\bmdefine{\biId}{I}
\bmdefine{\biJd}{J}
\bmdefine{\biKd}{K}
\bmdefine{\biLd}{L}
\bmdefine{\biMd}{M}
\bmdefine{\biOd}{N}
\bmdefine{\biPd}{O}
\bmdefine{\biQd}{P}
\bmdefine{\biRd}{R}
\bmdefine{\biSd}{S}
\bmdefine{\biTd}{T}
\bmdefine{\biUd}{U}
\bmdefine{\biVd}{V}
\bmdefine{\biWd}{W}
\bmdefine{\biXd}{X}
\bmdefine{\biYd}{Y}
\bmdefine{\biZd}{Z}

\bmdefine{\biDelta}{\Delta}
\bmdefine{\biLambda}{\Lambda}
\bmdefine{\biPhi}{\Phi}
\bmdefine{\biSigma}{\Sigma}
\bmdefine{\biOmega}{\Omega}
\bmdefine{\biTheta}{\Theta}

\safemath{\bimA}{\biAd}
\safemath{\bimB}{\biBd}
\safemath{\bimC}{\biCd}
\safemath{\bimD}{\biDd}
\safemath{\bimE}{\biEd}
\safemath{\bimF}{\biFd}
\safemath{\bimG}{\biGd}
\safemath{\bimH}{\biHd}
\safemath{\bimI}{\biId}
\safemath{\bimJ}{\biJd}
\safemath{\bimK}{\biKd}
\safemath{\bimL}{\biLd}
\safemath{\bimM}{\biMd}
\safemath{\bimN}{\biNd}
\safemath{\bimO}{\biOd}
\safemath{\bimP}{\biPd}
\safemath{\bimQ}{\biQd}
\safemath{\bimR}{\biRd}
\safemath{\bimS}{\biSd}
\safemath{\bimT}{\biTd}
\safemath{\bimU}{\biUd}
\safemath{\bimV}{\biVd}
\safemath{\bimW}{\biWd}
\safemath{\bimX}{\biXd}
\safemath{\bimY}{\biYd}
\safemath{\bimZ}{\biZd}

\safemath{\bimDelta}{\biDelta}
\safemath{\bimLambda}{\biLambda}
\safemath{\bimPhi}{\biPhi}
\safemath{\bimSigma}{\biSigma}
\safemath{\bimOmega}{\biOmega}
\safemath{\bimTheta}{\biTheta}

\safemath{\setA}{\mathcal{A}}
\safemath{\setB}{\mathcal{B}}
\safemath{\setC}{\mathcal{C}}
\safemath{\setD}{\mathcal{D}}
\safemath{\setE}{\mathcal{E}}
\safemath{\setF}{\mathcal{F}}
\safemath{\setG}{\mathcal{G}}
\safemath{\setH}{\mathcal{H}}
\safemath{\setI}{\mathcal{I}}
\safemath{\setJ}{\mathcal{J}}
\safemath{\setK}{\mathcal{K}}
\safemath{\setL}{\mathcal{L}}
\safemath{\setM}{\mathcal{M}}
\safemath{\setN}{\mathcal{N}}
\safemath{\setO}{\mathcal{O}}
\safemath{\setP}{\mathcal{P}}
\safemath{\setQ}{\mathcal{Q}}
\safemath{\setR}{\mathcal{R}}
\safemath{\setS}{\mathcal{S}}
\safemath{\setT}{\mathcal{T}}
\safemath{\setU}{\mathcal{U}}
\safemath{\setV}{\mathcal{V}}
\safemath{\setW}{\mathcal{W}}
\safemath{\setX}{\mathcal{X}}
\safemath{\setY}{\mathcal{Y}}
\safemath{\setZ}{\mathcal{Z}}
\safemath{\emptySet}{\varnothing}

\safemath{\colA}{\mathscr{A}}
\safemath{\colB}{\mathscr{B}}
\safemath{\colC}{\mathscr{C}}
\safemath{\colD}{\mathscr{D}}
\safemath{\colE}{\mathscr{E}}
\safemath{\colF}{\mathscr{F}}
\safemath{\colG}{\mathscr{G}}
\safemath{\colH}{\mathscr{H}}
\safemath{\colI}{\mathscr{I}}
\safemath{\colJ}{\mathscr{J}}
\safemath{\colK}{\mathscr{K}}
\safemath{\colL}{\mathscr{L}}
\safemath{\colM}{\mathscr{M}}
\safemath{\colN}{\mathscr{N}}
\safemath{\colO}{\mathscr{O}}
\safemath{\colP}{\mathscr{P}}
\safemath{\colQ}{\mathscr{Q}}
\safemath{\colR}{\mathscr{R}}
\safemath{\colS}{\mathscr{S}}
\safemath{\colT}{\mathscr{T}}
\safemath{\colU}{\mathscr{U}}
\safemath{\colV}{\mathscr{V}}
\safemath{\colW}{\mathscr{W}}
\safemath{\colX}{\mathscr{X}}
\safemath{\colY}{\mathscr{Y}}
\safemath{\colZ}{\mathscr{Z}}

\safemath{\opA}{\mathbb{A}}
\safemath{\opB}{\mathbb{B}}
\safemath{\opC}{\mathbb{C}}
\safemath{\opD}{\mathbb{D}}
\safemath{\opE}{\mathbb{E}}
\safemath{\opF}{\mathbb{F}}
\safemath{\opG}{\mathbb{G}}
\safemath{\opH}{\mathbb{H}}
\safemath{\opI}{\mathbb{I}}
\safemath{\opJ}{\mathbb{J}}
\safemath{\opK}{\mathbb{K}}
\safemath{\opL}{\mathbb{L}}
\safemath{\opM}{\mathbb{M}}
\safemath{\opN}{\mathbb{N}}
\safemath{\opO}{\mathbb{O}}
\safemath{\opP}{\mathbb{P}}
\safemath{\opQ}{\mathbb{Q}}
\safemath{\opR}{\mathbb{R}}
\safemath{\opS}{\mathbb{S}}
\safemath{\opT}{\mathbb{T}}
\safemath{\opU}{\mathbb{U}}
\safemath{\opV}{\mathbb{V}}
\safemath{\opW}{\mathbb{W}}
\safemath{\opX}{\mathbb{X}}
\safemath{\opY}{\mathbb{Y}}
\safemath{\opZ}{\mathbb{Z}}
\safemath{\opZero}{\mathbb{O}}
\safemath{\identityop}{\opI}

\safemath{\veca}{\bma}
\safemath{\vecb}{\bmb}
\safemath{\vecc}{\bmc}
\safemath{\vecd}{\bmd}
\safemath{\vece}{\bme}
\safemath{\vecf}{\bmf}
\safemath{\vecg}{\bmg}
\safemath{\vech}{\bmh}
\safemath{\veci}{\bmi}
\safemath{\vecj}{\bmj}
\safemath{\veck}{\bmk}
\safemath{\vecl}{\bml}
\safemath{\vecm}{\bmm}
\safemath{\vecn}{\bmn}
\safemath{\veco}{\bmo}
\safemath{\vecp}{\bmp}
\safemath{\vecq}{\bmq}
\safemath{\vecr}{\bmr}
\safemath{\vecs}{\bms}
\safemath{\vect}{\bmt}
\safemath{\vecu}{\bmu}
\safemath{\vecv}{\bmv}
\safemath{\vecw}{\bmw}
\safemath{\vecx}{\bmx}
\safemath{\vecy}{\bmy}
\safemath{\vecz}{\bmz}

\safemath{\veczero}{\bmzero}
\safemath{\vecone}{\bmone}
\safemath{\vecxi}{\bmxi}
\safemath{\veclambda}{\bmlambda}
\safemath{\vecmu}{\bmmu}
\safemath{\vectheta}{\bmtheta}
\safemath{\vecphi}{\bmphi}
\safemath{\vecdelta}{\bmdelta}

\safemath{\matA}{\bA}
\safemath{\matB}{\bB}
\safemath{\matC}{\bC}
\safemath{\matD}{\bD}
\safemath{\matE}{\bE}
\safemath{\matF}{\bF}
\safemath{\matG}{\bG}
\safemath{\matH}{\bH}
\safemath{\matI}{\bI}
\safemath{\matJ}{\bJ}
\safemath{\matK}{\bK}
\safemath{\matL}{\bL}
\safemath{\matM}{\bM}
\safemath{\matN}{\bN}
\safemath{\matO}{\bO}
\safemath{\matP}{\bP}
\safemath{\matQ}{\bQ}
\safemath{\matR}{\bR}
\safemath{\matS}{\bS}
\safemath{\matT}{\bT}
\safemath{\matU}{\bU}
\safemath{\matV}{\bV}
\safemath{\matW}{\bW}
\safemath{\matX}{\bX}
\safemath{\matY}{\bY}
\safemath{\matZ}{\bZ}
\safemath{\matzero}{\bmzero}

\safemath{\matDelta}{\bDelta}
\safemath{\matLambda}{\bLambda}
\safemath{\matPhi}{\bPhi}
\safemath{\matSigma}{\bSigma}
\safemath{\matOmega}{\bOmega}
\safemath{\matTheta}{\bTheta}

\safemath{\matidentity}{\matI}
\safemath{\matone}{\matO}

\safemath{\rnda}{A}
\safemath{\rndb}{B}
\safemath{\rndc}{C}
\safemath{\rndd}{D}
\safemath{\rnde}{E}
\safemath{\rndf}{F}
\safemath{\rndg}{G}
\safemath{\rndh}{H}
\safemath{\rndi}{I}
\safemath{\rndj}{J}
\safemath{\rndk}{K}
\safemath{\rndl}{L}
\safemath{\rndm}{M}
\safemath{\rndn}{N}
\safemath{\rndo}{O}
\safemath{\rndp}{P}
\safemath{\rndq}{Q}
\safemath{\rndr}{R}
\safemath{\rnds}{S}
\safemath{\rndt}{T}
\safemath{\rndu}{U}
\safemath{\rndv}{V}
\safemath{\rndw}{W}
\safemath{\rndx}{X}
\safemath{\rndy}{Y}
\safemath{\rndz}{Z}

\safemath{\rveca}{\bimA}
\safemath{\rvecb}{\bimB}
\safemath{\rvecc}{\bimC}
\safemath{\rvecd}{\bimD}
\safemath{\rvece}{\bimE}
\safemath{\rvecf}{\bimF}
\safemath{\rvecg}{\bimG}
\safemath{\rvech}{\bimH}
\safemath{\rveci}{\bimI}
\safemath{\rvecj}{\bimJ}
\safemath{\rveck}{\bimK}
\safemath{\rvecl}{\bimL}
\safemath{\rvecm}{\bimM}
\safemath{\rvecn}{\bimN}
\safemath{\rveco}{\bomO}
\safemath{\rvecp}{\bimP}
\safemath{\rvecq}{\bimQ}
\safemath{\rvecr}{\bimR}
\safemath{\rvecs}{\bimS}
\safemath{\rvect}{\bimT}
\safemath{\rvecu}{\bimU}
\safemath{\rvecv}{\bimV}
\safemath{\rvecw}{\bimW}
\safemath{\rvecx}{\bimX}
\safemath{\rvecy}{\bimY}
\safemath{\rvecz}{\bimZ}

\safemath{\rvecxi}{\bmxi}
\safemath{\rveclambda}{\bmlambda}
\safemath{\rvecmu}{\bmmu}
\safemath{\rvectheta}{\bmtheta}
\safemath{\rvecphi}{\bmphi}

\safemath{\rmatA}{\bimA}
\safemath{\rmatB}{\bimB}
\safemath{\rmatC}{\bimC}
\safemath{\rmatD}{\bimD}
\safemath{\rmatE}{\bimE}
\safemath{\rmatF}{\bimF}
\safemath{\rmatG}{\bimG}
\safemath{\rmatH}{\bimH}
\safemath{\rmatI}{\bimI}
\safemath{\rmatJ}{\bimJ}
\safemath{\rmatK}{\bimK}
\safemath{\rmatL}{\bimL}
\safemath{\rmatM}{\bimM}
\safemath{\rmatN}{\bimN}
\safemath{\rmatO}{\bimO}
\safemath{\rmatP}{\bimP}
\safemath{\rmatQ}{\bimQ}
\safemath{\rmatR}{\bimR}
\safemath{\rmatS}{\bimS}
\safemath{\rmatT}{\bimT}
\safemath{\rmatU}{\bimU}
\safemath{\rmatV}{\bimV}
\safemath{\rmatW}{\bimW}
\safemath{\rmatX}{\bimX}
\safemath{\rmatY}{\bimY}
\safemath{\rmatZ}{\bimZ}

\safemath{\rmatDelta}{\bimDelta}
\safemath{\rmatLambda}{\bimLambda}
\safemath{\rmatPhi}{\bimPhi}
\safemath{\rmatSigma}{\bimSigma}
\safemath{\rmatOmega}{\bimOmega}
\safemath{\rmatTheta}{\bimTheta}

\usepackage{amssymb}
\usepackage{amsfonts}
\usepackage{mathrsfs}
\usepackage{xspace}
\usepackage{bm}
\usepackage{fancyref}
\usepackage{textcomp}

\usepackage{multirow}
\usepackage{stmaryrd}

\newenvironment{textbmatrix}{	\setlength{\arraycolsep}{2.5pt}%
								\big[\begin{matrix}}{\end{matrix}\big]%
								\raisebox{0.08ex}{\vphantom{M}}}

\def\be{\begin{equation}}
\def\ee{\end{equation}}
\def\een{\nonumber \end{equation}}
\def\mat{\begin{bmatrix}}
\def\emat{\end{bmatrix}}
\def\btm{\begin{textbmatrix}}
\def\etm{\end{textbmatrix}}

\def\ba#1\ea{\begin{align}#1\end{align}}
\def\bas#1\eas{\begin{align*}#1\end{align*}}
\def\bs#1\es{\begin{split}#1\end{split}}
\def\bg#1\eg{\begin{gather}#1\end{gather}}
\def\bml#1\eml{\begin{multline}#1\end{multline}}
\def\bi#1\ei{\begin{itemize}#1\end{itemize}}

\newcommand{\lefto}{\mathopen{}\left}

\DeclareMathOperator{\Exop}{\opE}			

\newcommand{\Ex}[2]{\ensuremath{\Exop_{#1}\lefto[#2\right]}} 	

\safemath{\dirac}{\delta}					
\safemath{\krond}{\dirac}					

\safemath{\upto}{\uparrow}
\safemath{\downto}{\downarrow}
\safemath{\iu}{j}							
\safemath{\ev}{\lambda}						
\safemath{\hilseqspace}{l^{2}}				
\newcommand{\banachfunspace}[1]{\setL^{#1}}	
\safemath{\hilfunspace}{\banachfunspace{2}}	

\safemath{\SNR}{\textit{SNR}} 				
\safemath{\PAR}{\textit{PAR}} 				
\safemath{\No}{N_0}							
\safemath{\Es}{E_s}							
\safemath{\Eb}{E_b}							
\safemath{\EbNo}{\frac{\Eb}{\No}}
\safemath{\EsNo}{\frac{\Es}{\No}}

\DeclareMathOperator{\CHop}{\ensuremath{\opH}} 
\safemath{\tvir}{\rndh_{\CHop}}				
\safemath{\tvtf}{\rndl_{\CHop}}				
\safemath{\spf}{\rnds_{\CHop}}				
\safemath{\bff}{H_{\CHop}}					

\safemath{\ircf}{r_{h}}						
\safemath{\tftvcf}{r_{s}}					
\safemath{\tfcf}{r_{l}}						
\safemath{\bfcf}{r_{H}}						

\safemath{\tcorr}{c_h}						
\safemath{\scf}{c_{s}}						
\safemath{\tfcorr}{c_{l}}					
\safemath{\fcorr}{c_{H}}						

\safemath{\mi}{I}							
\safemath{\capacity}{C}						

\safemath{\normal}{\mathcal{N}}			
\safemath{\jpg}{\mathcal{CN}}			
\safemath{\mchain}{\leftrightarrow}		

\safemath{\dB}{\,\mathrm{dB}}
\safemath{\dBm}{\,\mathrm{dBm}}
\safemath{\Hz}{\,\mathrm{Hz}}
\safemath{\kHz}{\,\mathrm{kHz}}
\safemath{\MHz}{\,\mathrm{MHz}}
\safemath{\GHz}{\,\mathrm{GHz}}
\safemath{\s}{\,\mathrm{s}}
\safemath{\ms}{\,\mathrm{ms}}
\safemath{\mus}{\,\mathrm{\text{\textmu}s}}
\safemath{\ns}{\,\mathrm{ns}}
\safemath{\ps}{\,\mathrm{ps}}
\safemath{\meter}{\,\mathrm{m}}
\safemath{\mm}{\,\mathrm{mm}}
\safemath{\cm}{\,\mathrm{cm}}
\safemath{\m}{\,\mathrm{m}}
\safemath{\W}{\,\mathrm{W}}
\safemath{\mW}{\, \mathrm{mW}}
\safemath{\J}{\,\mathrm{J}}
\safemath{\K}{\,\mathrm{K}}
\safemath{\bit}{\,\mathrm{bit}}
\safemath{\nat}{\,\mathrm{nat}}

\safemath{\define}{\triangleq}			

\safemath{\equivalent}{\sim}
\safemath{\distas}{\sim}					
\safemath{\sdiff}{\Delta}				

\safemath{\reals}{\mathbb{R}}
\safemath{\positivereals}{\reals_{+}}
\safemath{\integers}{\mathbb{Z}}
\safemath{\posint}{\integers_{+}}
\safemath{\naturals}{\mathbb{N}}
\safemath{\posnaturals}{\naturals_{+}}
\safemath{\complexset}{\mathbb{C}}
\safemath{\rationals}{\mathbb{Q}}

\newcommand*{\fancyrefapplabelprefix}{app}		
\newcommand*{\fancyrefthmlabelprefix}{thm}		
\newcommand*{\fancyreflemlabelprefix}{lem}		
\newcommand*{\fancyrefcorlabelprefix}{cor}		
\newcommand*{\fancyrefdeflabelprefix}{def}		
\newcommand*{\fancyrefproplabelprefix}{prop}		
\newcommand*{\fancyrefexmpllabelprefix}{exmpl}
\newcommand*{\fancyrefalglabelprefix}{alg}		
\newcommand*{\fancyreftbllabelprefix}{tbl}		

\frefformat{vario}{\fancyrefseclabelprefix}{Section~#1}
\frefformat{vario}{\fancyrefthmlabelprefix}{Theorem~#1}
\frefformat{vario}{\fancyreftbllabelprefix}{Table~#1}
\frefformat{vario}{\fancyreflemlabelprefix}{Lemma~#1}
\frefformat{vario}{\fancyrefcorlabelprefix}{Corollary~#1}
\frefformat{vario}{\fancyrefdeflabelprefix}{Definition~#1}
\frefformat{vario}{\fancyreffiglabelprefix}{Fig.~#1}
\frefformat{vario}{\fancyrefapplabelprefix}{Appendix~#1}
\frefformat{vario}{\fancyrefeqlabelprefix}{(#1)}
\frefformat{vario}{\fancyrefproplabelprefix}{Proposition~#1}
\frefformat{vario}{\fancyrefexmpllabelprefix}{Example~#1}
\frefformat{vario}{\fancyrefalglabelprefix}{Algorithm~#1}

\safemath{\dictab}{[\,\dicta\,\,\dictb\,]}

\safemath{\ysig}{\bmy}
\safemath{\ysighat}{\hat{\ysig}}
\safemath{\ysigdim}{M}
\safemath{\xsig}{\bmx}
\safemath{\xsigdim}{N}
\safemath{\nx}{n_x}
\safemath{\zsig}{\bmz}
\safemath{\zsigdim}{\ysigdim}
\safemath{\rsig}{\bmr}
\safemath{\Adict}{\bA}
\safemath{\Adicttilde}{\widetilde{\Adict}}
\safemath{\Adictdim}{\outputdim\times\xsigdim}
\safemath{\avec}{\bma}
\safemath{\avectilde}{\tilde{\avec}}
\safemath{\Bdict}{\bB}
\safemath{\Bdicttilde}{\widetilde{\Bdict}}
\safemath{\Cdict}{\bC}
\safemath{\cvec}{\bmc}
\safemath{\Ddict}{\bD}
\safemath{\Ddictdim}{\ysigdim\times\xsigdim}
\safemath{\dvec}{\bmd}
\safemath{\Ddicttilde}{\widetilde{\bD}}
\safemath{\Bonb}{\bB}
\safemath{\bvec}{\bmb}
\safemath{\Bonbdim}{\ysigdim\times\ysigdim}
\safemath{\noise}{\bmn}
\safemath{\noisedim}{\ysigim}
\safemath{\err}{\bme}
\safemath{\errdim}{\ysigdim}
\safemath{\errset}{\setE}
\safemath{\nerr}{n_e}
\safemath{\delop}{\bP_\errset}
\safemath{\delopc}{\bP_{{\errset}^c}}

\safemath{\cplxi}{\imath}
\safemath{\cplxj}{\jmath}

\safemath{\dict}{\matD}
\safemath{\inputdim}{N}		
\safemath{\outputdim}{M}		
\safemath{\sparsity}{S}	
\safemath{\inputdimA}{{N_a}}	
\safemath{\inputdimB}{{N_b}}	
\safemath{\elemA}{{n_a}}	
\safemath{\elemB}{{n_b}}	
\safemath{\resA}{\matR_a}	
\safemath{\resB}{\matR_b}	
\safemath{\subD}{\matS} 
\safemath{\subA}{\matS_a} 
\safemath{\subB}{\matS_b} 
\safemath{\dicta}{\matA} 	
\safemath{\dictb}{\matB} 	
\safemath{\hollowS}{H}
\safemath{\hollowA}{H_a}
\safemath{\hollowB}{H_b}
\safemath{\cross}{Z}
\safemath{\coh}{\mu_d}			
\safemath{\coha}{\mu_a}			
\safemath{\cohb}{\mu_b}			
\safemath{\mubs}{\nu}	
\safemath{\cohm}{\mu_m} 
\safemath{\dictset}{\setD}	
\safemath{\dictsetp}{\dictset(\coh,\coha,\cohb)}	
\safemath{\dictsetgen}{\dictset_\text{gen}}
\safemath{\dictsetgenp}{\dictsetgen(\coh)}
\safemath{\dictsetonb}{\dictset_\text{onb}}
\safemath{\dictsetonbp}{\dictsetonb(\coh)}

\safemath{\leftside}{U}
\safemath{\rightsideA}{R_a}
\safemath{\rightsideB}{R_b}

\safemath{\indexS}{\setI_S} 

\safemath{\na}{n_a}			
\safemath{\nb}{n_b}			
\safemath{\coeffa}{p_i}	
\safemath{\coeffb}{q_j}	
\safemath{\seta}{\setP}		
\safemath{\setb}{\setQ}     
\safemath{\setw}{\setW}	
\safemath{\setz}{\setZ}	
\safemath{\cola}{\veca}		
\safemath{\colb}{\vecb}		
\safemath{\cold}{\vecd}		
\safemath{\inputvec}{\vecx} 	
\safemath{\error}{\vece}	
\safemath{\noiseout}{\vecz} 	
\safemath{\inputvecel}{x}
\safemath{\inputveca}{\vecx_a}
\safemath{\inputvecb}{\vecx_b}
\safemath{\outputvec}{\vecy}	
\safemath{\lambdamin}{\lambda_{\mathrm{min}}}

\safemath{\elltwo}{\ell_2}
\safemath{\ellone}{\ell_1}
\safemath{\ellzero}{\ell_0}
\safemath{\ellinf}{\ell_\infty}
\safemath{\ellinftilde}{\ell_{\widetilde\infty}}
\safemath{\licard}{Z(\coh,\coha,\cohb)}
\safemath{\xsol}{\hat{x}}
\safemath{\xbord}{x_b}		
\safemath{\xstat}{x_s}		
\safemath{\xstatLone}{\tilde{x}_s}
\safemath{\order}{\mathcal{O}} 
\safemath{\scales}{\Theta} 
\safemath{\ones}{\mathbf{1}} 
\safemath{\zeroes}{\mathbf{0}} 
\safemath{\thlone}{\kappa(\coh,\cohb)} 
\safemath{\constoneA}{\delta} 
\safemath{\constoneB}{\epsilon} 
\safemath{\nlarge}{L}				   
\safemath{\sumlarge}{S_\nlarge}
\safemath{\maxlarger}{P_\nlarge}	   
\safemath{\Pzero}{\textrm{P0}}	
\safemath{\Pone}{\textrm{P1}}
\safemath{\vecfir}{\vecw}			 
\safemath{\vecsec}{\vecz}
\safemath{\elvecfir}{w}              
\safemath{\elvecsec}{z}				 
\safemath{\nlargefir}{n}
\safemath{\normout}{\gamma}
\safemath{\auxfun}{h}
\safemath{\supp}{\textrm{supp}}

\safemath{\indexa}{\ell}
\safemath{\indexb}{r}
\safemath{\indexc}{i}
\safemath{\indexd}{j}

\safemath{\project}{P}

\begin{document}

\title{Analog vs. Digital Spatial Transforms: \\ A Throughput, Power, and Area Comparison
}

 \author{\IEEEauthorblockN{Zephan M. Enciso$^\text{1}$, Seyed Hadi Mirfarshbafan$^\text{2}$, Oscar Casta\~neda$^\text{2}$, \\ Clemens JS. Schaefer$^\text{1}$, Christoph Studer$^\text{3}$, and  Siddharth Joshi$^\text{1}$}\\
 
   \IEEEauthorblockA{$^\text{1}$Department of Computer Science and Engineering, University of Notre Dame, Notre Dame, IN, USA } 
  \IEEEauthorblockA{$^\text{2}$Department of Electrical and Computer Engineering, Cornell Tech, New York, NY, USA} 
 \IEEEauthorblockA{$^\text{3}$Department of Information Technology and Electrical Engineering, ETH Zurich, Zurich, Switzerland} 
  e-mail: {\em zenciso@nd.edu; sm2675@cornell.edu; oc66@cornell.edu; cschaef6@nd.edu; studer@ethz.ch; sjoshi2@nd.edu}
\thanks{The work of SHM, OC, and CS was supported by ComSenTer, one of six centers in JUMP, a Semiconductor Research Corporation (SRC) program sponsored by DARPA. The work of OC and CS was also supported by Xilinx, Inc.\ and by the US NSF under grants ECCS-1408006, CCF-1535897,  CCF-1652065, CNS-1717559, and ECCS-1824379. SJ was supported in part by NSF/Intel Partnership on Machine Learning for Wireless Networking Systems under grant CNS-2002921.}

 }


\maketitle

\begin{abstract}
Spatial linear transforms that process multiple parallel analog signals to simplify downstream signal processing find widespread use in multi-antenna communication systems, machine learning inference, data compression, audio and ultrasound applications, among many others. 
In the past, a wide range of mixed-signal as well as digital spatial transform circuits have been proposed---it is, however, a longstanding question whether analog or digital transforms are superior in terms of throughput, power, and area.
In this paper, we focus on Hadamard transforms and perform a systematic comparison of state-of-the-art analog and digital circuits implementing spatial transforms in the same 65\,nm CMOS technology. 
We analyze the trade-offs between throughput, power, and area, and we identify regimes in which mixed-signal or digital Hadamard transforms are preferable.
Our comparison reveals that (i) there is no clear winner and (ii) analog-to-digital conversion is often dominating area and energy efficiency---and not the spatial transform. 
\end{abstract}

\IEEEpeerreviewmaketitle

\section{Introduction and Contributions}

Sensing and processing multiple analog signal channels simultaneously is commonly encountered in a variety of fields including healthcare (ultrasound), multi-antenna communication, machine learning, imaging, and computer vision. Efficiently processing parallel streams of analog signals remains a challenging task due to the increasingly stringent latency and energy requirements imposed on the underlying hardware. Because spatial transforms, in contrast to spectral or time-interleaved transforms, have no temporal dependencies between inputs, they are highly amenable to parallel processing in area and energy efficient analog and digital circuits. This property of spatial transforms naturally raises the question of whether spatial transforms are more efficiently implemented using analog circuitry or through digital designs.

Previous work~\cite{joshi2017algorithms} indicates that analog spatial processing can be efficiently implemented using capacitor arrays. These results suggest that analog processing prior to digitization can relax the requirements of the analog-to-digital converters (ADCs), improving the system's overall energy efficiency. 
Digital transforms come in various flavors, including streaming and time-interleaved architectures; see, e.g.,  \cite{puschel2005spiral}. However, not much is known about the efficacy of massively-parallel transforms that are suitable for spatial processing of high-dimensional signals.
Most importantly, to the best of our knowledge, no systematic comparison between analog and digital spatial transforms exists, which leaves the question of which of the two approaches is more beneficial in practice. 

This paper represents a first attempt to systematically compare state-of-the-art analog and digital circuit designs with respect to area, throughput, and power for implementing spatial transforms. 
We focus on analog and digital circuits for spatial Hadamard transforms implemented in the same commercial, general-purpose 65\,nm CMOS technology. 
We first detail the analog and digital circuit designs, provide reference post-layout implementation results, and compare their input and output signal-to-noise ratio (SNR) behaviors. 
We then study the area efficiency (area per throughput) and energy efficiency (power per throughput) trade-offs by considering the area and power of ADCs.
Our comparison enables us to identify operation regimes for which analog or digital designs are preferable.

\section{Background}

\subsection{Hadamard Transform Basics}
\begin{figure}[tp]
\centering
\includegraphics[width=0.5\columnwidth]{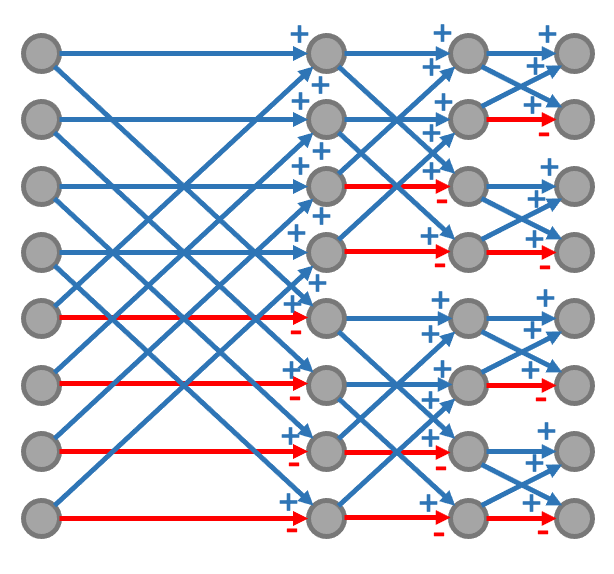}
\vspace{-0.25cm}
\caption{Illustration of the dataflow graph of an $M=8$ fast Hadamard transform (FHT). The FHT consists of $m=\log_2(M)=3$ stages each performing $M/2$ two-dimensional Hadamard transforms on permuted inputs.}
\label{fig:FHT.png}
\end{figure}

In order to compare analog vs. digital spatial transforms, we focus on the Hadamard transform (HT), which finds widespread use for data compression, compressive sensing,  imaging, and locality sensitive hashing. 
The Hadamard transform is essentially a matrix-vector product of a Hadamard matrix  $\bH_m$ by a vector $\bmx\in\reals^M$ with $M=2^m$, i.e., $\bmy=\bH_m\bmx$.
A Hadamard matrix~$\bH_m$ of dimension $2^m\times2^m$ can be constructed recursively. By defining $\bH_0=1$, we can construct  Hadamard matrices for natural numbers $m$ as 
\begin{align}
\label{eq:hadamard}
\bH_m = \frac{1}{\sqrt{2}} \left[
\begin{array}{cc}
+\bH_{m-1} & +\bH_{m-1} \\
+\bH_{m-1} & -\bH_{m-1}
\end{array}\right].
\end{align}
To avoid an explicit matrix-vector product that involves \mbox{$M^2-M$} additions and subtractions, one typically resorts to the fast Hadamard transform (FHT). 
The FHT repeatedly applies $2^{m-1}$ Hadamard transforms of size $m=2$ (so-called radix-2 butterfly operations $\bmy=\bH_2\bmx$) in $m$ stages as illustrated by the dataflow graph in~\fref{fig:FHT.png}. Note the scale factors $1/\sqrt{2}$, which ensure that Euclidean norms are preserved, i.e.,  $\|\bmy\|=\|\bmx\|$, can be compensated either in every stage or at the end of the FHT; for the explicit Hadamard transform, the scale factors are typically included at the end of the matrix-vector product.
The digital Hadamard transform implementation relies on the FHT, whereas the analog Hadamard transform effectively implements an explicit matrix-vector product using only capacitors. 

\subsection{Prior Analog/Mixed-Signal Spatial Transform}
The analog circuit implementing HT closely follows the principles developed in previously fabricated mixed-signal spatial signal processing circuits~\cite{joshi:2017:isscc}. This prototype implements analog matrix-vector multiplication using continuous-time multiplying digital-to-analog converters (MDACs) to form the matrix coefficients, which are then multiplied with differential analog inputs. Using capacitors in this fashion results in highly linear circuits that (i) weight the analog AC signals and (ii) linearly sum them onto a common node, resulting in $84$\,dB of signal separation performance for real-time beamforming of multiple-input multiple-output (MIMO) orthogonal frequency-division multiplexing (OFDM) signals~\cite{joshi:2017:isscc}. Each capacitor in the MDAC uses a shielded structure in which, driven bottom and top plates shield the internal node from parasitics. By implementing continuous-time weighting of the analog signal, one mitigates capacitor switching and thus minimizes both $CV^2$ switching energy and $kT/C$ noise. Consequently, capacitor sizing is primarily determined by matching requirements.
We will describe a suitable analog HT design in \fref{sec:ms-circuit}.

\subsection{Prior Digital Spatial Transform}

The fast Fourier transform (FFT) is among the most prominent digital spatial transforms and finds widespread use in communication systems, e.g., for beamspace processing \cite{mirfarshbafan2020sparse}.
FFT hardware design is an extremely mature area and state-of-the-art FFT designs can be generated automatically with SPIRAL \cite{puschel2005spiral}.
In contrast, only a handful of custom FHT designs have been reported in the open literature; see, e.g.,~\cite{huang2005analysis}.
Existing parallel FHTs support relatively small dimensions (e.g. up to $M=16$) and are typically applied to two-dimensional images for data compression.
FHTs are extremely hardware-friendly as they only involve additions and subtractions. Furthermore, the simplicity of in-place processing minimizes the storage of sequential HT engines. Nevertheless, not much is known for larger Hadamard transforms that are suitable for spatial processing. 
We will describe a digital FHT design suitable for spatial processing in \fref{sec:dig-circuit}.
\vspace{-.25cm}

\section{Implementation Details }\label{sec:implement}

\subsection{Mixed-Signal Implementation}\label{sec:ms-circuit}
\begin{figure}[tp]
\centering
\includegraphics[width=0.85\columnwidth]{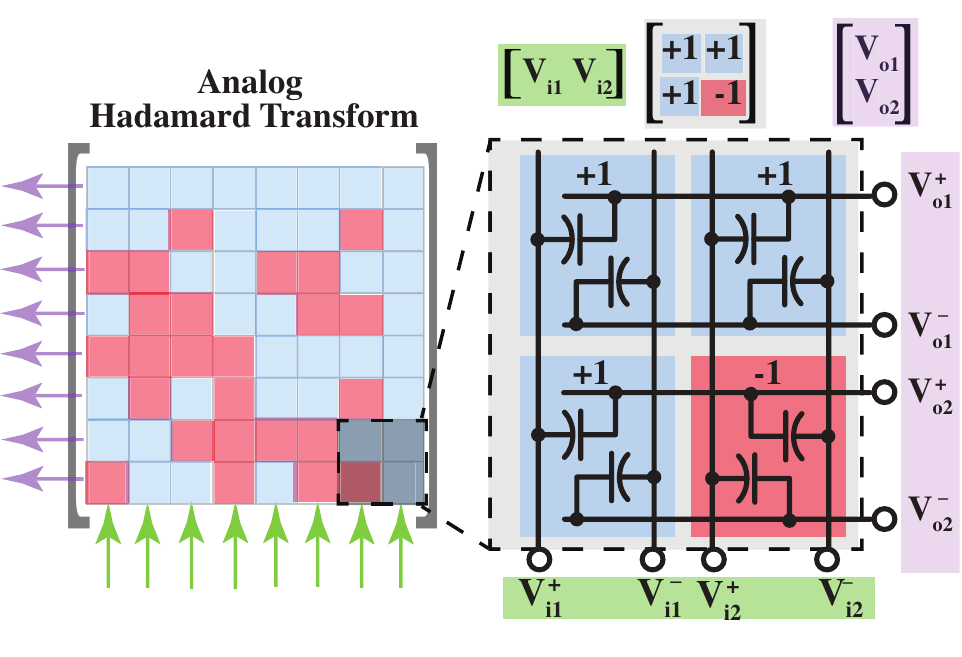}
\vspace{-0.25cm}
\caption{Illustration of an $8\times8$ Hadamard transform matrix with details provided for a representative $2\times2$ sub-block. Differentially encoded inputs, $V^+_{ij}$ and $V^-_{ij}$, are either added or subtracted onto an output differential pair, $V^+_\text{ok}$ and$V^-_\text{ok}$, through capacitive coupling. The addition/subtraction occurs when the Hadamard transform matrix entry is a $+1$/$-1$.}
\label{fig:ms-circuit}
\vspace{-0.0cm}
\end{figure}

Our analog HT implements a $128\times 128$ HT matrix using a differential capacitor structure as shown in~\fref{fig:ms-circuit}. The inputs and the outputs of this block are continuous-time, differential analog signals. Since the HT is a fixed-transform, this leads to a compact cell which is then repeatedly tiled in layout, each bottom plate is driven by one of the polarities of the differential signals. We place two complementary instances of the array to ensure that both polarities of the signal see a constant capacitive load.
The capacitor array was laid out in TSMC 65\,nm CMOS, with the capacitors occupying metal layers 4, 5, and 6. The area, maximum frequency, and unit capacitor values entered in~\fref{tbl:analogresults} are from post-extraction simulation, they were verified against 10b to 14b data converters~\cite{joshi:2017:isscc,joshi:jssc:2016} that we previously taped out. To derive a realistic cut-off frequency for the system, we set the output conductance  of the array drivers to 14\,$\mu$S. 
\fref{tbl:analogresults} summarizes the design across multiple array sizes. When C$_\text{unit}=4$\,fF the HT capacitor array size is comparable to the digital implementations in~\fref{tbl:digitalresults}. Aggressive scaling of the unit capacitors to sub-femto-farad results in f$_{3\,\text{dB}} > 4$\,GHz and consequently a f$_{\,\text{nyq.}} > 8$\,GHz.

\begin{table}[tp]
  \caption{Post-layout results for 128-point analog Hadamard transforms with different unit capacitors in 65\,nm CMOS}
  \label{tbl:analogresults}
  \centering
  \begin{tabular}{@{}lcccc@{}}
  \toprule
C$_\text{unit}$ & C$_\text{unit}$ area & Array area & f$_{3\text{dB}}$@14.4\,$\mu$S   & Cap. mismatch \\

[fF] & [$\mu$m$^\text{2}$] & [mm$^\text{2}$] & g$_\text{driver}$ [GHz] & $\sigma_u$/C$_u$ [arb.~unit]\\
\midrule
0.68&  2.25&  0.078  & 4.65  & 0.06\\
1.5 & 4.41 &  0.153  &  2.55 & 0.024\\
2.0 & 5.76 &  0.200  &   2.03& 0.016\\
4.0 & 10.24 & 0.356  &   1.1 & 0.01\\
    \bottomrule
  \end{tabular}
  \vspace{-0.3cm}
\end{table}

\subsection{Digital Architecture and Implementation}\label{sec:dig-circuit}

Our digital FHT implements a fully-unrolled decimation in frequency architecture using radix-2 butterflies, as illustrated in \fref{fig:FHT.png}. 
The 128-point FHT implementation consists of $m=7$ stages, where each stage contains $64$ radix-2 butterflies that perform addition and subtraction of the two inputs.
Since the output bitwidth of an adder/subtractor is one bit more than that of its input, we  allow the odd-numbered stages to increase the bitwidth by one---the even-numbered stages apply a scale factor of $\frac{1}{2}$, thereby maintaining the bitwidth. Consequently, the outputs of the design have only 4b more resolution than the inputs, which reduces area and ensures proper normalization of the FHT.

In order to minimize the critical path of our FHT design, the outputs of each stage are pipelined.

\fref{tbl:digitalresults} shows post-layout results for 128-point FHTs ranging from 5b to 10b input precision in TSMC 65\,nm CMOS. 
We note that these are---to the best of our knowledge---the first implementation results of digital 128-point Hadamard transforms reported in the open literature. 
The cell density is around 80\% for all digital designs. Since our architecture is fully unrolled and pipelined, the maximum sustained throughput (in transforms per second) equals the maximum clock frequency. 
The area and net power consumption scale roughly linearly with the number of input bits and the precision has a marginal effect on the maximum clock frequency. 

\begin{table}[tp]
  \caption{Post-layout results for 128-point digital fast Hadamard transforms (FHTs) with 5b to 10b input precision in 65\,nm CMOS}
  \label{tbl:digitalresults}
  \centering
  \begin{tabular}{@{}lccccc@{}}
  \toprule
Input res. & Area & Max.~freq. & Power &   Area eff.\ & Energy eff.\  \\

[bit] & [mm$^\text{2}$] & [GHz] & [mW] & [mm$^\text{2}$/GT/s] & [pJ/T] \\
\midrule
5 & 0.195 &    1.603 & 346.7 & 0.122 & 216.4 \\
6 & 0.236 &    1.605 & 431.4 &  0.147 & 268.8 \\
7 & 0.277 &    1.439 & 440.6 &   0.192 & 306.2 \\
8 & 0.314 &  1.429  & 517.0 &   0.219 & 361.9 \\
9 & 0.341 &   1.431 & 575.9 & 0.239 & 402.5 \\
10 & 0.394 &    1.377  & 617.1 &  0.287 & 448.0 \\
    \bottomrule
  \end{tabular}
\end{table}

\section{Comparison}\label{sec:comparison}
\subsection{Methodology}
\begin{figure}[tp]
\centering
\subfigure[Analog transform]{\includegraphics[width=0.475\columnwidth]{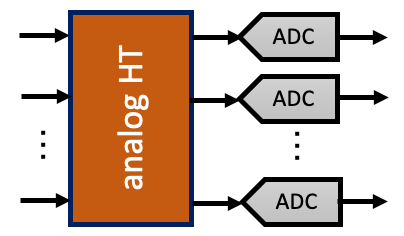}}
\hspace{0.1cm}
\subfigure[Digital transform]{\includegraphics[width=0.475\columnwidth]{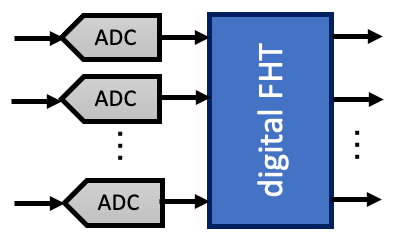}}
\caption{Comparison methodology. For the analog transform, we first apply the Hadamard transform using passive, capacitor circuits followed by converting the analog signal using 128 ADCs; for the digital transform, we first use 128 ADCs followed by the digital fast Hadamard transform (FHT).}
\label{fig:comp.png}
\end{figure}

\begin{figure}
\centering
\subfigure[Analog transform]{\includegraphics[width=0.492\columnwidth]{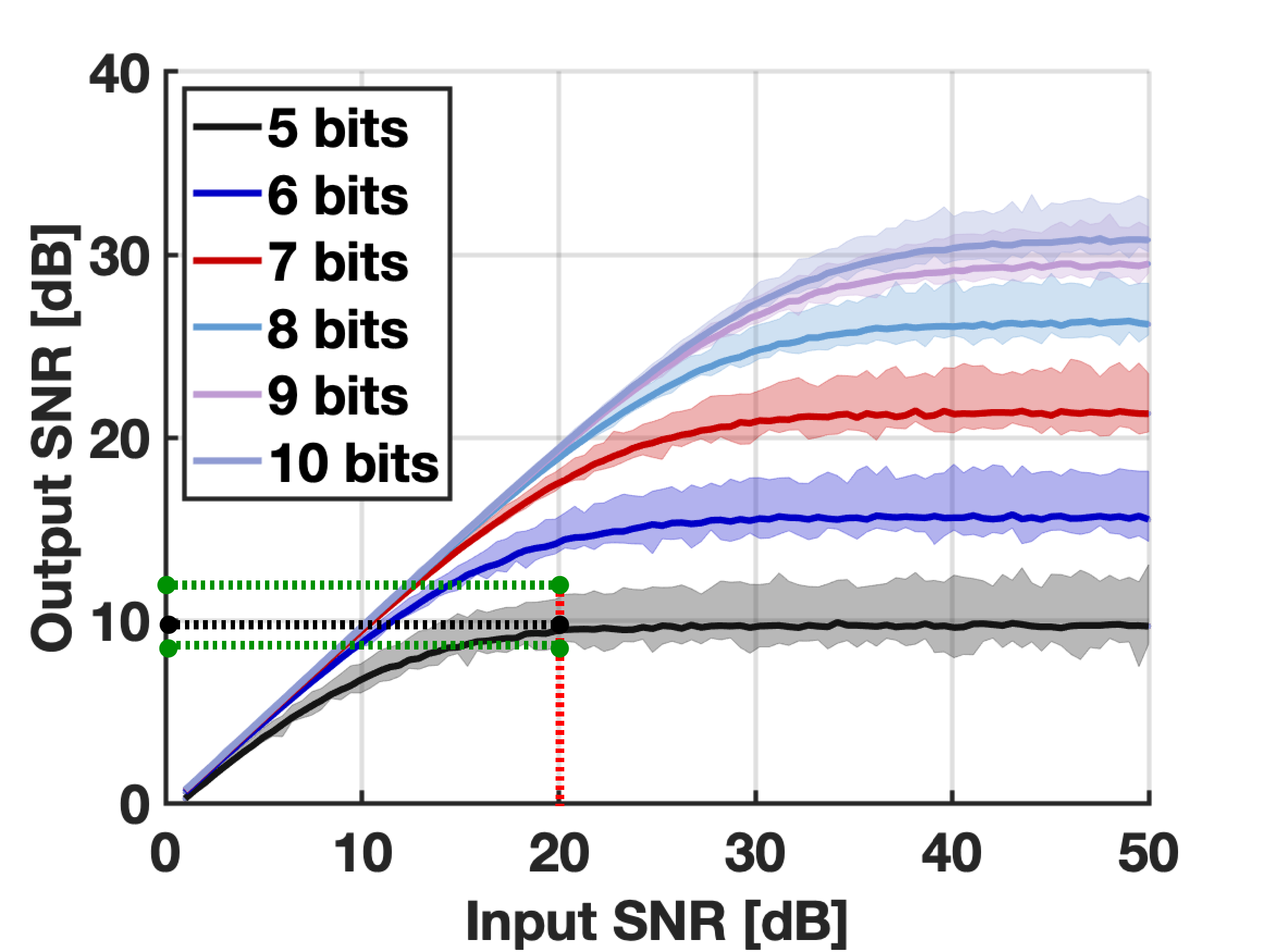}\label{fig:snrsmismatch}}
\subfigure[Digital transform]{\includegraphics[width=0.492\columnwidth]{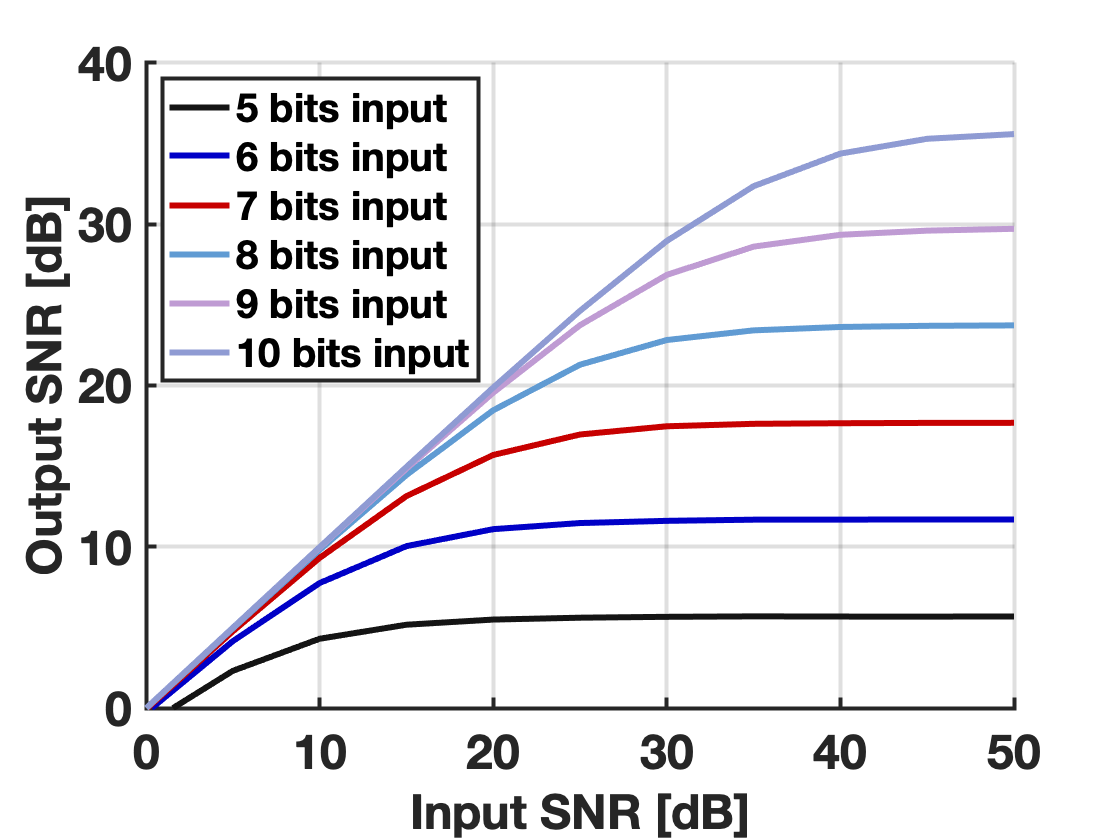}\label{fig:snrsweep:digital}}
\caption{Input vs.\ output SNR for analog and digital Hadamard transforms. (a) Shows the effect of quantization and capacitor mismatch for the analog HT implemented using a capacitor array composed of $0.68$\,fF unit capacitors. The shaded area represents the spread in achievable output SNR with a solid line, representing the lowest point for a 90\% yield. At an input SNR of $20$\,dB the spread in output SNR due to mismatch is highlighted by the dotted lines. (b) Shows the output precision of the digital FHT design.}
\end{figure}\vspace{-.2cm}

\begin{figure*}[t]
\centering
\subfigure[Energy efficiency\label{fig:powerefficiency}]{\includegraphics[width=0.325\textwidth]{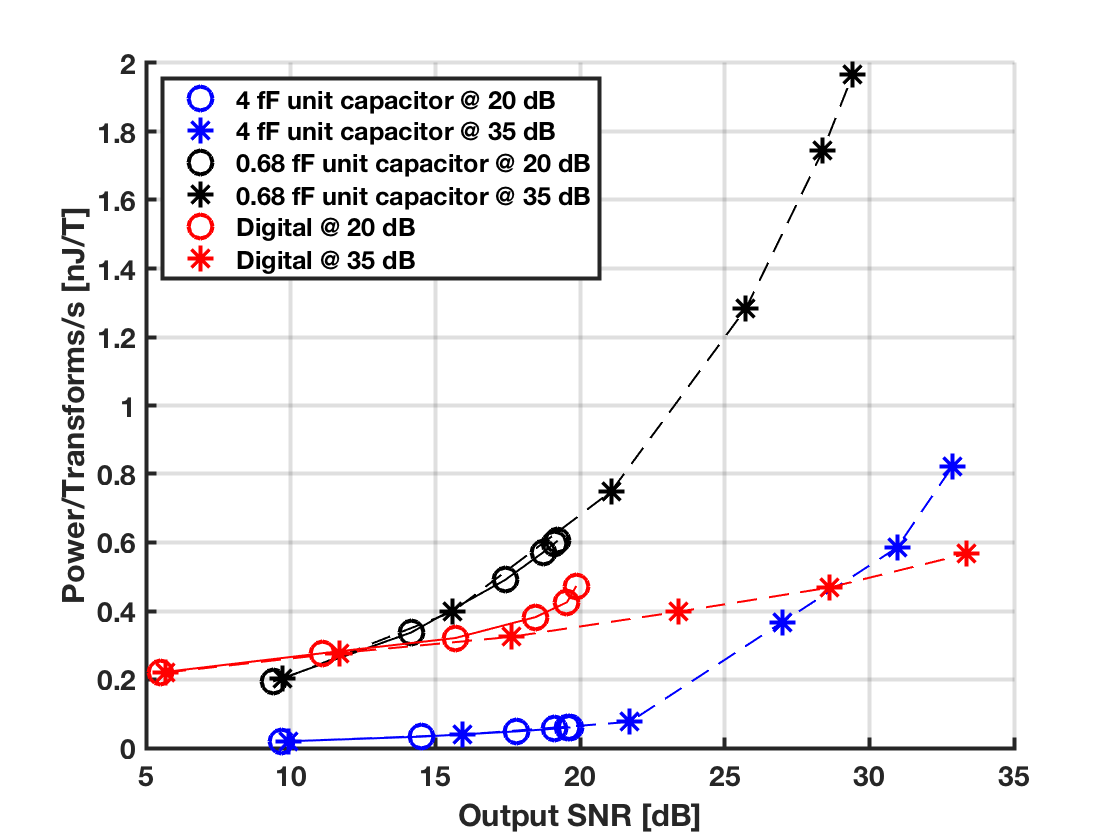}}
\subfigure[Area efficiency excluding ADCs\label{fig:throughput:noadc}]{\includegraphics[width=0.325\textwidth]{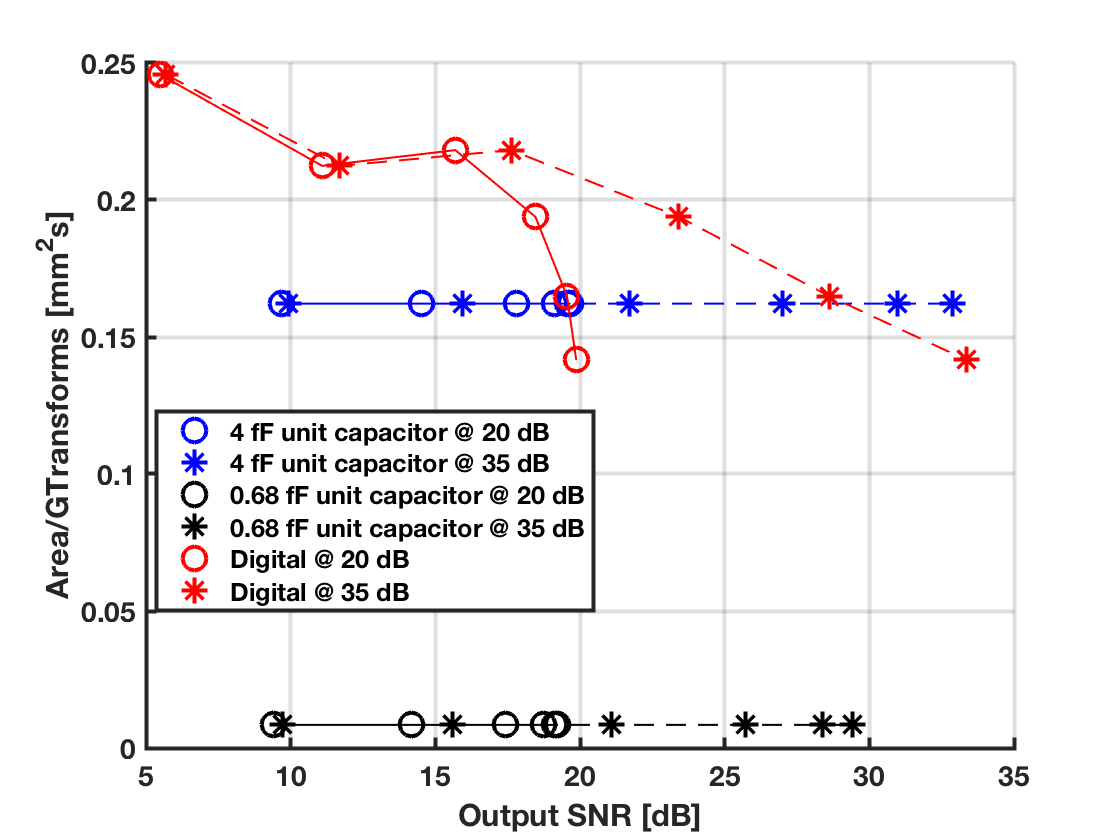}}
\subfigure[Area efficiency including ADCs\label{fig:throughput:adc}]{\includegraphics[width=0.325\textwidth]{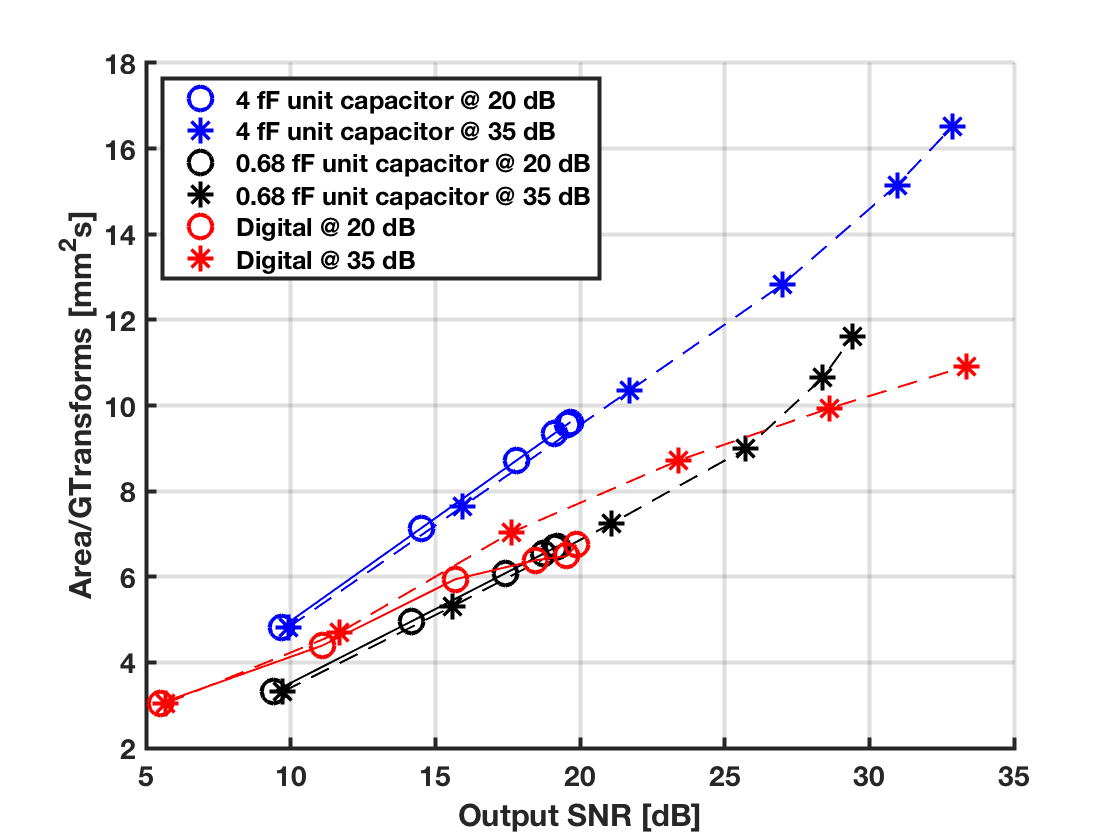}}
\caption{Energy and area efficiency vs.\ output SNR trade-offs.
(a) Although the analog design with $0.68$\,fF unit capacitors achieves higher $f_{3\text{dB}}$, operating at such frequencies requires expensive ADCs, which annihilate the benefit of  compact analog circuitry. 
The analog design with $4$\,fF unit unit capacitors achieves lower $f_{3\text{dB}}$, which is conducive to power efficient ADCs. %
For the digital FHT, the ADC power is comparable to that of the digital part.
(b) Shows  the area efficiency without the ADC area, which reveals that analog transforms can be more compact and suffer from no area increase due to the fixed array size.
(c) Shows the area efficiency with the ADC area, which shows that the ADC area is substantial, effectively resulting in designs of comparable efficiency.}
\label{fig:results.png}\vspace{-.3cm}
\end{figure*}

\fref{fig:comp.png} illustrates the comparison methodology used in this paper. 
In order to arrive at a fair comparison between both approaches, we include the area and power of  analog-to-digital converters (ADCs)  that would otherwise be present in a real-world system. Additionally we account for signal attenuation incurred during the analog transform ($11.3$\,dB for the $128$-point HT), by correspondingly increasing the SNR requirement from the downstream ADC.
To this end, for the analog transform, we first use the analog Hadamard transform design detailed in \fref{sec:ms-circuit} followed by a dedicated ADC for each of the $128$ analog outputs. For the digital transform, we first use a set of $128$ ADCs to convert the analog inputs followed by the digital FHT design. 
For both transform designs, we pick ADCs from~\cite{murmann2020adc} that match the resolution with signal-to-quantization-noise ratio (SQR) of the analog or digital transform, as well as the maximum achievable bandwidth by the individual designs.

\subsection{Input SNR vs. Output SNR}
As a first step, we study the accuracy and linearity of the two approaches. 
To characterize the input and output SNR, we consider the input signal model $\bmx=\bms+\bmn$, where $\bms$ is the signal vector and $\bmn$ is the noise vector; both are i.i.d. zero-mean Gaussian. The signal and noise variances are determined by input SNR. 
We then measure the output SNR as
\begin{align}
\textit{SNR}_\text{out} = \frac{\Ex{}{\|\bmy\|^2}}{\Ex{}{\|\bmy-\hat{\bmy}\|^2}},
\end{align}
where $\bmy=\bH\bms$ is the output of an ideal, noise-free Hadamard transform and $\hat{\bmy}$ is the quantized output of transforming $\bmx=\bms+\bmn$ using either the analog HT or the digital FHT. 

For the analog design, we consider the effect of capacitor mismatch on the HT. All analog HT results were extracted from 400 Monte--Carlo trials of capacitor mismatch with 400 trials per SNR.
Using the methodology described in~\cite{omran:2016:matchingfemtofarad} and our fabricated IC~\cite{joshi:2017:isscc}, we estimate the mismatch coefficient for the capacitors to be $A=2\%\sqrt{1\,\text{fF}}$.

\fref{fig:snrsmismatch} shows the effect of this mismatch for $C_\text{unit}=0.68$\,fF on the SNR of a transformed signal, for various output ADC resolutions. At a target input SNR of 20\,dB, the mismatch creates a spread of possible values; the dotted lines in \fref{fig:snrsmismatch} indicate the maximum and minimum output SNRs observed over 400 Monte--Carlo trials for an input SNR of 20\,dB.
For the digital transform, we use a bit-true golden model to extract the output SNR via Monte--Carlo simulations. \fref{fig:snrsweep:digital} shows the SNR transfer behavior of the digital FHT. We observe that the output SNR is lower than that of the analog transform for less than 7b input resolution---for higher resolution, the digital FHT achieves higher output SNR.

\subsection{Area-efficiency and Energy-efficiency Trade-offs}

\fref{fig:powerefficiency} compares the energy efficiency obtained from two analog configurations (with unit capacitors $4$\,fF and $0.68$\,fF) and the digital implementations. While the analog HT design with the smaller unit capacitor operates at a higher bandwidth, the energy and area overheads of high-frequency ADCs are detrimental to the combined system efficiency. Indeed, the $4$\,fF array shows superior energy efficiency than the $0.68$\,fF array, primarily due to a more energy-efficient ADC.
As expected, at higher resolutions (output SNR $\ge 30$\,dB), the digital design is more energy-efficient. Examining the energy contribution of the ADCs shows that the ADC power is comparable to the power of the digital FHT power, but it dominates the power of the analog HT. This disparity is explained by the ADC SNDR increasing by $12$~dB to compensate for capacitor induced attentuation in the analog signal path (insertion loss). 

\fref{fig:throughput:noadc} compares the area efficiency of the three designs, where we exclude the ADC area. In this comparison, the analog circuits are much more area efficient, with the smaller array ($C_\text{unit}=0.68$\,fF) delivering an order of magnitude higher throughput than the digital FHT. 
However, when ADC area is included in the comparison, \fref{fig:throughput:adc} reveals that this advantage is immediately negated. Indeed, the area efficiency for all three designs now becomes comparable, in part due to the costly ADCs required for high-speed operation. 
Moreover, we cannot identify a clear design point that is better across categories, i.e., while the slower operation due to larger capacitors leads to improved energy efficiency, the larger area also reduces throughput. As expected, the digital FHT is consistently better than analog HTs at very high resolution---when ADC overheads are completely accounted for.

\section{Conclusions and Outlook}\label{sec:conclusion}

We studied the area and energy efficiency of implementing spatial Hadamard transforms through passive analog circuits and massively-parallel digital circuits. All of our designs have been implemented in the same 65\,nm CMOS technology. 
Our analysis reveals that neither design is an outright winner in all categories.
We note that the Hadamard transform uniquely advantages the analog design, leading to extremely compact and energy-efficient implementations. Despite this, our analysis reveals that the ADCs heavily influence the overall area and energy efficiency of spatial Hadamard transforms, indicating that further optimizations must include data converter design. For analog spatial transforms to truly deliver, we would need: (i) the ADC  to be co-designed with the analog processing and (ii) circuit topologies that exploit transform sparsity must be employed to minimize insertion loss.
Finally, an extensive comparison between analog and digital spatial Fourier transforms, which are useful for emerging millimeter-wave communications systems, is part of future work. \vspace{-.3cm}

\bibliographystyle{IEEEtran}
\bibliography{mwcas} 
\vspace{-.2cm}

\end{document}